# Holographic photonic neuron

Vincent R. Daria[*]

*Research School of Physics, Australian National University, Canberra, Australia*
*Eccles Institute of Neuroscience, John Curtin School of Medical Research, Australian National University, Canberra, Australia*

The promise of artificial intelligence (AI) to process complex datasets has brought about innovative computing paradigms. While recent developments in quantum-photonic computing have reached significant feats, mimicking our brain's ability to recognize images are poorly integrated in these ventures. Here, I incorporate orbital angular momentum (OAM) states in a classical Vander Lugt optical correlator to create the holographic photonic neuron. The photonic neuron can memorize an array of matched filters in a phase-hologram, which is derived by linking OAM states with elements in the array. Successful correlation is independent of intensity and yields photons with OAM states of $l\hbar$, which can be used as a transmission protocol or qudits for quantum computing. The unique OAM identifier establishes the photonic neuron as a fundamental AI device for pattern recognition that can be scaled and integrated with other computing platforms to build-up a neuromorphic quantum-photonic processor that mimics the brain.

**I. Introduction**

Rapid technological advancements have led to an exponential growth of portable devices setting up the stage for a fully interconnected world. Alongside these developments comes the need to process the associated large and complex datasets, which requires new computing paradigms. Silicon-based neuromorphic or biologically inspired artificial intelligence (AI) technologies allow for more effective deep learning algorithms capable of handling complex and large datasets[1-3]. However, neuromorphic photonics have now emerged as a promising option[4-6]. A neuromorphic silicon-photonic processor has been shown to solve the Lorenz attractor differential equation[7]. While the computing speed has improved, storing and readout of optical memories have difficulty coping up with clock speeds in gigahertz frequencies[8-10]. The bottleneck also occurs when transferring information data from one computing platform to the other[11], which hinders scalability and integration.

Yet processing clock speeds may not be all that matter. There is a large disparity when comparing AI technologies with how our brain works. Apart from the large gap in energy consumption, the temporal dynamics of synaptic processes in our brain operate in kilohertz range and still outperform computing functions, such as machine vision. With our current understanding of memory formation[12,13] and vision processing[14,15] in the brain, our own experiences can tell that while we can instantaneously view and process highly resolved details of visual scenes, some of these details are not fully stored in our brain. Processing of visual data are concatenated into pertinent information and stored as synaptic strengths in networks of neurons in different regions of our brain. When recalling our memories, our brain offers seamless transfer of information from different brain regions making it possible to recognize patterns in a probabilistic manner. Hence, if we were to get inspiration from this model, building an intelligent neuromorphic computer implies a system with high resolution instantaneous processing, concatenated storage of pertinent information, distributed processing and the ability to seamlessly integrate and transmit information within multiple platforms such as deterministic with probabilistic computing operations.

There has been significant work on optical implementations of machine vision systems based on the Vander Lugt optical correlator[16,17]. The correlator epitomizes classical optical computing that measures the similarity between two patterns. The optical correlator is implemented via a 4*f* lens setup where the multiplication of a pre-determined matched filter and the Fourier transform of the input pattern is performed at the Fourier plane[18]. Matched filters in a Vander Lugt correlator represent memory storage that could provide instantaneous readout and processing at the speed of light. Matched filters can also be applied using only its phase[19] and can be multiplexed and stored as phase holograms[20-23].

---

[*] vincent.daria@anu.edu.au

Over 50 years have passed and the Vander Lugt correlator remains an effective tool for pattern recognition problems such alignment of Deoxyribonucleic acid (DNA) sequences[24] or counting k-mers substrings in a DNA string[25]. Pattern recognition can also be achieved via optical artificial neural networks (ANNs)[26,27]. Diffractive deep neural networks (D²NNs) use multiple holographic layers, where each point in the hologram acts as a neuron with a complex-valued transmission coefficient[28]. To arrive at a desired optical field at the output, Fresnel light propagation through the layers intricately incorporates lens functions within the holograms during learning. On the other hand, the Vander Lugt correlator employs classical optical computing, where Fourier transform operations are performed by lenses that effectively sets up light paths like a fully connected ANN. One could therefore argue that the Vander Lugt correlator is a 3-layer ANN, where the 1st and 3rd layers take the form of diffractive lenses. Training such 3-layer ANN entails finding the appropriate complex-valued transmission of the 2nd layer, which is essentially the matched filter. And while D²NNs can use more than 3 layers, probabilistic pattern recognition is still performed via classical computing operations.

Despite these recent advances in neuromorphic[6] and optical AI technologies[29], probabilistic classical optical computing operations are not yet fully integrated with quantum computing. Recent investments indicate that quantum technologies will be the computing platform for future AI implementations[30]. While a quantum optical neural network has been proposed to perform a number of quantum information processing tasks[31], pattern recognition was not among them. In an effort to demonstrate a quantum-based Vander Lugt correlator, Qiu, et al. [32] used a pump beam with a Laguerre-Gaussian (LG) mode in a ghost imaging setup. To perform quantum correlation, the interaction between the spatial frequency information of the object and the filter is performed via down-converted photon pairs at the Fourier domain. The orbital angular momentum (OAM) based quantum Vander Lugt correlator, however, was only demonstrated to perform vortex mapping and identification[32].

Here, I propose a generalized Vander Lugt correlator that can integrate classical optical computing with quantum computing. I refer this technique as the holographic photonic neuron, which produces output photons carrying OAM states of $l\hbar$. Aside from circumventing intensity dependence, the OAM of the LG laser modes links macroscopic classical optics with quantum effects[33] especially with the realization of OAM-entangled photon pairs[34-36] and pairs of simultaneously entangled OAM and spin angular momentum for superdense coding in quantum communications[37]. Multiplexed OAM states have also been used as photonic qubits in quantum computers[38] and as multi-level qudits in quantum information protocols[39]. By integrating classical computing operations with quantum-optics[31], quantum technologies[40] and silicon-photonics[6,7], we can realize a robust platform for neuromorphic quantum photonics.

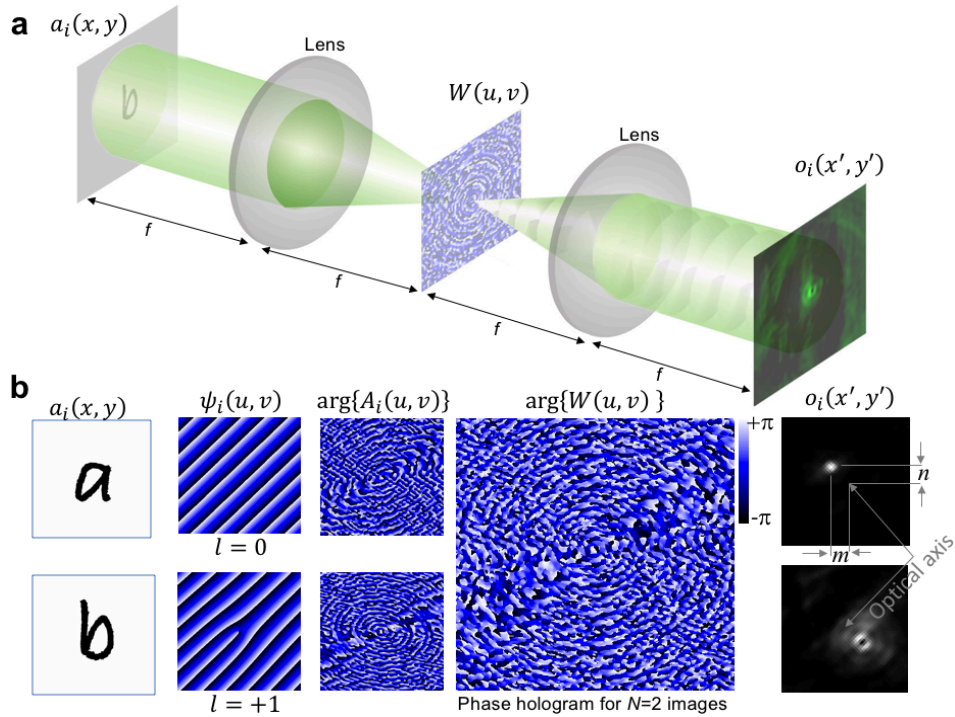

**Figure 1. The holographic photonic neuron.** (a) Classical Vander Lugt optical correlator with a phase-only matched filter that yields photons with orbital angular momentum (OAM) states as unique identifier. (b) Illustration of operation where an input pattern, $a_i(x,y)$, whose Fourier transform, $A_i(u,v)$, is linked with a carrier phase function, $\psi_i(u,v)$, and linearly combined with other patterns to form $W_i(u,v)$. The carrier phase function can be encoded with an optical vortex to produce output photons with OAM of topological charge $l$. The top row shows an input pattern "𝑎" is encoded with $l=0$, which yields a Gaussian correlation peak, while the bottom row shows an input pattern "𝑏" is encoded with $l=+1$ producing an optical vortex.

## II. Results

**Holographic photonic neuron**

The holographic photonic neuron is implemented via a 4$f$ lens setup with a multiplexed matched filter at the Fourier plane (**Figure 1a**). The multiplexed matched filter contains information from a training set consisting of an array of visual data. The matched filter of an $i^{th}$ element in an array of $N$ elements is related to the two-dimensional (2D) Fourier transform of an input pattern $a_i(x, y)$ given by,

$$A_i(u, v) = \mathcal{F}\{a_i(x, y)\} \quad (1)$$

where $(x, y)$ are spatial coordinates at the input plane and $(u, v)$ are spatial frequency coordinates at the Fourier plane. Training the holographic photonic neuron links spatially distinct, non-overlapping, LG modes to individual elements in the array via a carrier phase function, $\psi_i(u, v)$, given by

$$\psi_i(u, v) = 2\pi(m \cdot u + n \cdot v) + l\varphi \quad (2)$$

where $\varphi = \arctan\{v/u\}$ is the azimuthal angle around the optical axis, $l$ is the topological charge of an LG beam, and $(m, n)$ are the spatial locations of the correlation peaks at the output. By linear superposition[20,21], the resultant field of $N$ matched filters, is given by

$$W(u, v) = \sum_{i=0}^{N} \exp\big(j(\psi_i(u, v) - \arg\{A_i(u, v)\})\big) \quad (3)$$

whose 2D phase profile $\arg\{W(u, v)\}$ is the hologram containing pertinent information of $a_i(x, y)$. On readout, the output field of the optical correlation is given by

$$o_i(x', y') = \mathcal{F}^{-1}\{\exp(\arg\{W(u, v)\}) \cdot A_i(u, v)\} \quad (4)$$

**Figure 1b**, graphically describes the process for a training set of $N$=2 patterns ("*a*" and "*b*"). The carrier phase functions, $\psi_i(u, v)$, with OAM states $l$=0 and

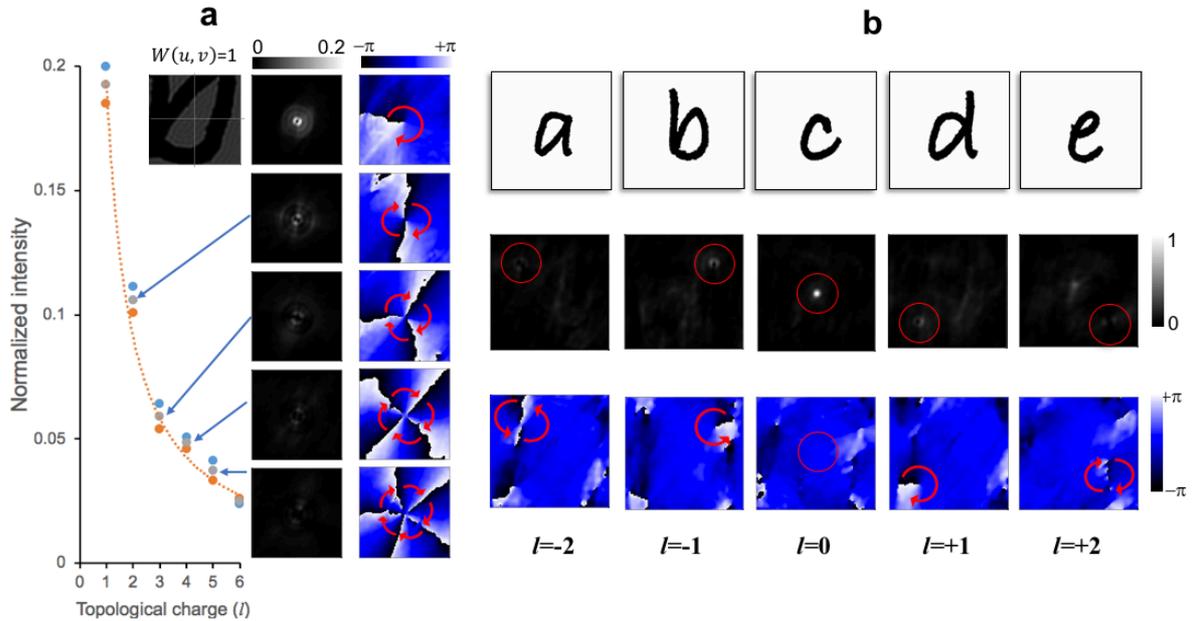

**Figure 2. Optical pattern recognition with phase-singularities at the output.** (**a**) The correlation peak intensity resulting from the autocorrelation of a pattern "*a*" encoded with OAM states from $l$=+1 to $l$=+6. The 2D intensity and phase distributions are shown per $l$ setting. Red curved arrows indicate the direction of helicity with phase values from -π to +π. (**b**) The output intensity and phase distributions resulting from the cross-correlation of patterns "*a*" to "*e*" encoded with OAM from $l$=-2 to $l$= +2, respectively.

$l$=+1 (with $m=n$=constant) are linked to patterns "a" and "b", respectively. Consequently, $W(u,v)$ is derived for $N$=2 matched filters. The positions ($m, n$) of the correlation peaks at the output plane can be set arbitrarily, which will depend on the succeeding stage of computing or transmission such as an optical fiber bundle or a multi-channel detector.

**Encoding OAM states**

The performance of the holographic photonic neuron is verified via numerical simulations and a benchtop experimental setup. Details for the numerical simulation and experiments are discussed in the methods section. **Figure 2a** plots the correlation peak intensity resulting from the autocorrelation of the pattern "a" with OAM states from $l$=+1 to $l$= +6. Autocorrelation of an input pattern is performed when $W(u,v)$ is derived with the matched filter of the same pattern ($N$=1). The correlation peak position is set at the optical axis ($m=n$=0) and the intensity is normalized to the maximum intensity when $l$=0. The intensity along the vortex is not constant and the three points plotted per OAM state are the maximum, minimum and average intensities. The zoomed-in images of the vortices and their corresponding phase patterns are shown. For reference, the amount of zoom can be compared with the conjugate image of the input pattern "a" when $W(u,v) = 1$. While the maximum intensity of the correlation peak is drastically reduced to noise level at high $l$, the phase-singularity and the OAM states at the output can still be detected.

**Figure 2b** shows the correlation of letters "a" to "e" (Bradley Hand font) using a $W(u,v)$ derived with $N$=5 matched filters. The intensity and phase distributions at the output for OAM states from $l$=-2 to $l$=+2 are shown. The spatial positions of the correlation peaks (red circles) are arranged symmetrically along the optical axis ($m=n$=0) where the correlation peak for pattern "c" ($l$=0) is located. The phase-singularities can be discerned from the phase distributions, which can be detected using spiral imaging[41] or quantum-based vortex mapping [32,42].

Experimental demonstration is performed using a green laser ($\lambda$=532 nm) and a phase-only spatial light modulator (SLM) (**Figure 3a**). The input patterns are printed on a transparency and imaged through the 4$f$ imaging setup where the SLM is placed at the Fourier plane and a camera at the output. Prior to performing optical correlation, $W(u,v)$ was derived with no OAM ($l$=0) by calculating the inverse Fourier transform of the image of the input pattern "a" (Arial font). The image of the input pattern is acquired by turning the SLM off, which effectively sets $W(u,v) = 1$ (**Figure 3b**). When the SLM is turned on and encoded with arg{$W(u,v)$}, the autocorrelation of the input pattern "a" yields a Gaussian correlation peak ($l=0$) (**Figure 3d**). Next, I calculated $W(u,v)$ for $N=1$ but with different OAM states from $l=-3$ to $l = +3$. **Figure 3e** shows optical vortices corresponding to the OAM states resulting from the autocorrelation of the pattern "a". A $W(u,v)$ with $N=2$ matched filters for patterns "a" and "b" can also be calculated with different OAM states. **Figure 3f** shows the correlation peaks for $l=0$, while **Figure 3g** shows the output for two different OAM states of $l=-1$ and $l=+1$ linked to letters "a" and "b", respectively.

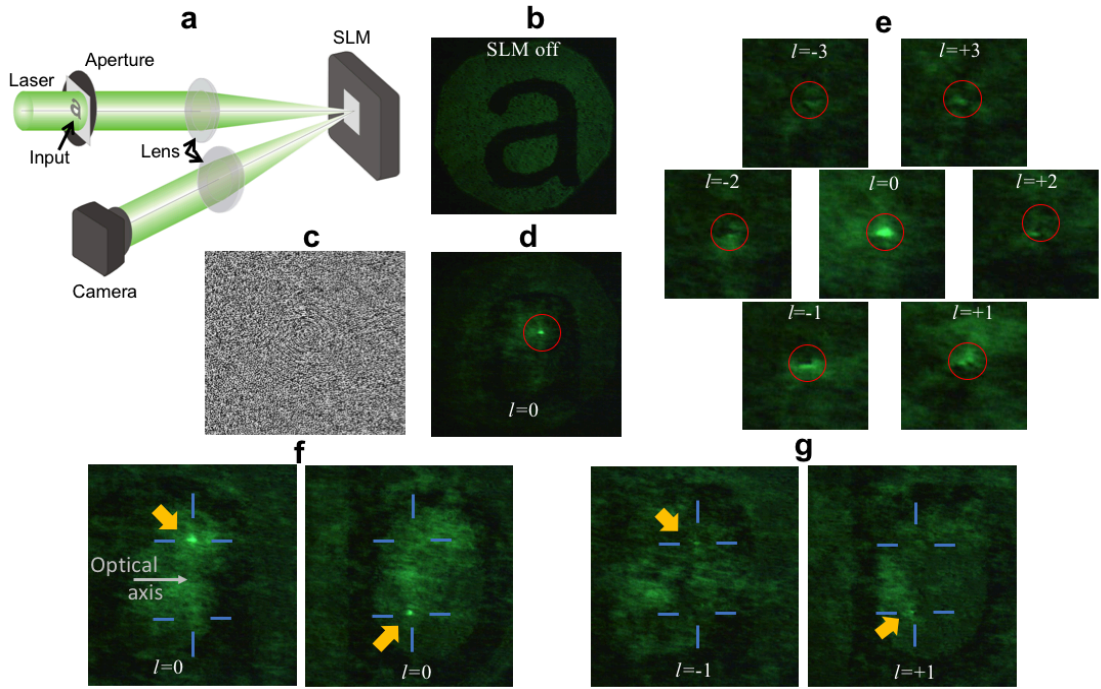

**Figure 3. Experimental verification of the holographic photonic neuron.** (**a**) Optical setup of the Vander Lugt correlator where arg$\{W(u,v)\}$ is encoded on a phase-only SLM located at the Fourier plane. (**b**) Output image of the input pattern when $W(u,v) = 1$ or when the SLM is turned off. (**c**) Representative phase hologram (arg$\{W(u,v)\}$) derived with $N$=1 matched filter and no OAM ($l$=0). (**d**) The autocorrelation of "a" yields a Gaussian correlation peak. (**e**) The autocorrelation of "a" yields an output where the OAM is varied from $l$=-3 to $l$=+3. The correlation of letters "a" and "b" using a training set with $N$=2 matched filters and with OAM states (**f**) $l$=0 and (**g**) $l$=-1 for "a" and $l$=+1 for "b".

**Intensity and noise dependence**

The OAM at the output is independent of the performance of the holographic photonic neuron. Hence, we can assess the correlation efficacy using $l=0$, where it represents a Vander Lugt correlator with multiplexed phase-only matched filters. **Figure 4a** is data from the experiment using a training set with $N=2$ patterns, "a" and "b". Setting the input pattern to either "a" or "b" results in a correlation peak (yellow arrows). Since the input patterns are switched and moved along the *x*-axis, the correlation peaks are assigned with non-overlapping *y*-axis positions. On the other hand, input patterns "c" and "d", which are not within the training set, did not yield correlation peaks.

Increasing the number of multiplexed matched filters yields lower correlation peak intensities. Using numerical simulation, complex patterns from 36 face photographs[43] and 36 alphanumeric characters (several fonts) were used to assess the correlation and discrimination efficacy. Color photographs were first converted to grayscale before performing correlation. For easier analysis, the positions of the correlation peaks at the output are arranged in a square pattern (e.g. $N=4=2\times2$, $N=9=3\times3$, up to $N=36=6\times6$). **Figure 4b** shows correlation results using a $W(u,v)$ with $N=9$ matched filters for a training set containing normal contrast (NC) face photographs, while **Figure 4c** shows results for $N=16$ alphanumeric characters (Bradley Hand font). As with **Figure 4a**, the top two patterns in **Figures 4b** and **4c** are patterns within the training set yielding correlation peaks (yellow arrows), while the bottom patterns are not. Note that in **Figure 4b** and **4c**, similarities in the NC face photographs (NC#15 and NC#01) and patterns "m" and "w" yield false positives (red circles). I will discuss this issue in the next section, together with **Figures 5** and **6**.

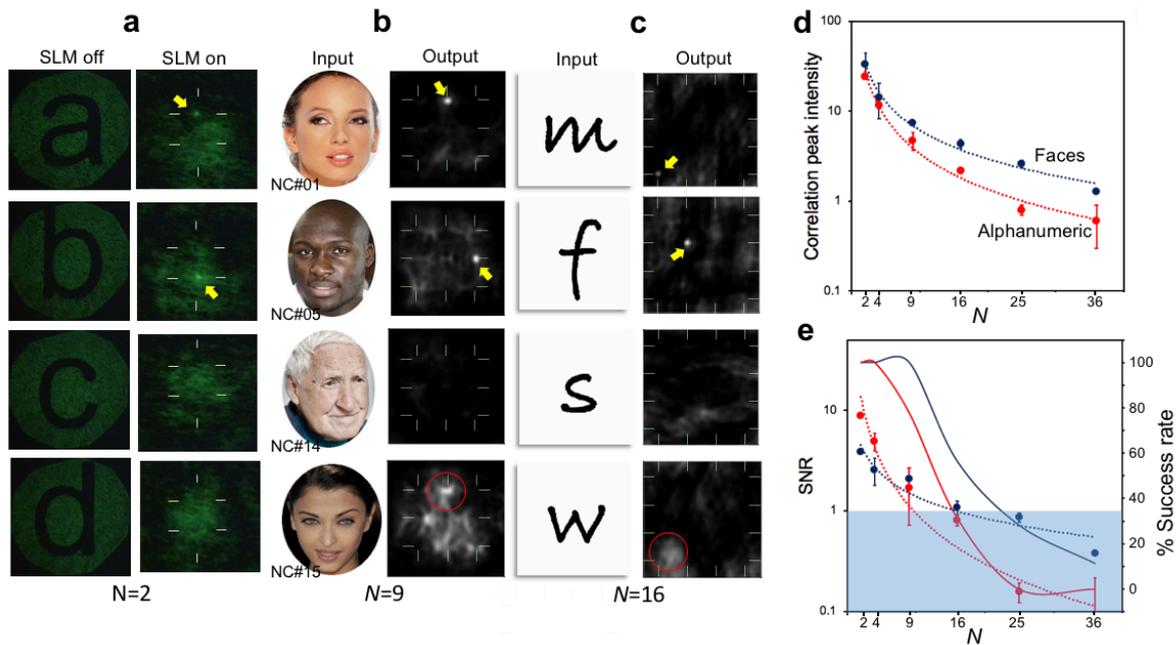

**Figure 4. Correlation efficacy of the multiplexed matched filters. (a - c)** The optical correlation performance of the holographic photonic neuron using $W(u,v)$ with (**a**) $N=2$, (**b**) $N=9$, and (**c**) $N=16$ matched filters. The top two patterns are included in the training set yielding correlation peaks indicated in yellow arrows, while the bottom two patterns are not included in the training set. (**d**) Plot of the correlation peak intensity as a function of $N$. (**e**) Plot of the SNR (left *y*-axis, dashed marked traces) and success rate (right *y*-axis, solid unmarked traces) for alphanumeric (red trace) and face photographs (blue trace) as a function of $N$. Peak intensities within the blue shaded area are below the noise level.

On average, the correlation peak intensities for face photographs and alphanumeric characters follow a $1/N$ profile as expected from the conservation of energy (**Figure 4d**). The linear combination of $N$ matched filters is equivalent to producing holographically projected multiple foci from a single laser[44]. However, the output of the holographic photonic neuron produces only a single correlation peak, which projects the rest of the photons as background noise. Hence, it is important to consider the signal-to-noise ratio (SNR) of the correlation peaks when detecting peak intensities.

The SNR is determined by the ratio of the correlation peak intensity over the peak-to-peak noise (**Supplementary Figure 1**). **Figure 4e** plots the SNR (left axis, marked traces) for alphanumeric patterns (red trace) and face photographs (blue trace) as a function $N$. There is a stark difference between high-contrast simple patterns such as alphanumeric characters (red dashed trace) over complex multi-level images of faces (blue dashed trace). Compared to face photographs, the SNR for alphanumeric characters is relatively high for $N < 9$, but drastically degrades at higher $N$. The blue shaded area indicates when the correlation peak is less than the noise. **Figure 4e** also plots the success rate for identifying correlation peaks (right axis, unmarked traces) for alphanumeric patterns (red trace) and face photographs (blue trace). The success rate is the percentage ratio of the number of correlation peaks with SNR > 1. For $N=9$, the success rate for identifying faces is still 100% (blue solid trace), while alphanumeric characters (red solid trace) have large deviations with success rate of 77.8%.

**False negatives and false positives**

Large deviations in correlation peak intensities yield false negatives, where the peak intensity is below the background noise. **Figure 5a** shows indistinguishable peaks for patterns "𝑖" and "𝑙" using a $W(u,v)$ with $N=16$ alphanumeric patterns (Bradley hand font). High correlation peaks for "𝑗" and "𝑘" with intensity profile (gray traces) along the $x'$-axis are shown for comparison. On the other hand, the intensity profiles show that the correlation peaks for "𝑖" (blue trace) and "𝑙" (red trace) are below the background noise. However, when the matched filters are linked with OAM state $l=+1$, the output yields a helical phase profile at the location of the correlation peak. Hence, the phase singularities where $l \neq 0$ can still be detected even when the correlation peak with $l=0$ is below the noise level.

Similarities in the patterns yield a secondary correlation peak, which can reduce the efficacy of identification. **Figure 5b** shows secondary peaks resulting from the correlation of patterns "c" and "e" using $W(u,v)$ with $N$=9 alphanumeric characters (Times New Roman font). Moreover, using a carrier phase function with OAM states $l$=+1 and $l$=-1 for patterns "c" and "e", respectively, also returns opposing helical phase singularities in both 1st and 2nd correlation peaks, respectively. Hence, detecting phase singularities to differentiate the two characters will not be effective. While the true positive peak is higher than the 2nd peak, these issues are common problems in machine vision[45]. Approaches based on convolutional neural networks to decompose the spatial frequencies of the pattern into sub-groups can be used to optimize discrimination between similar patterns. I will touch on this issue in the discussion.

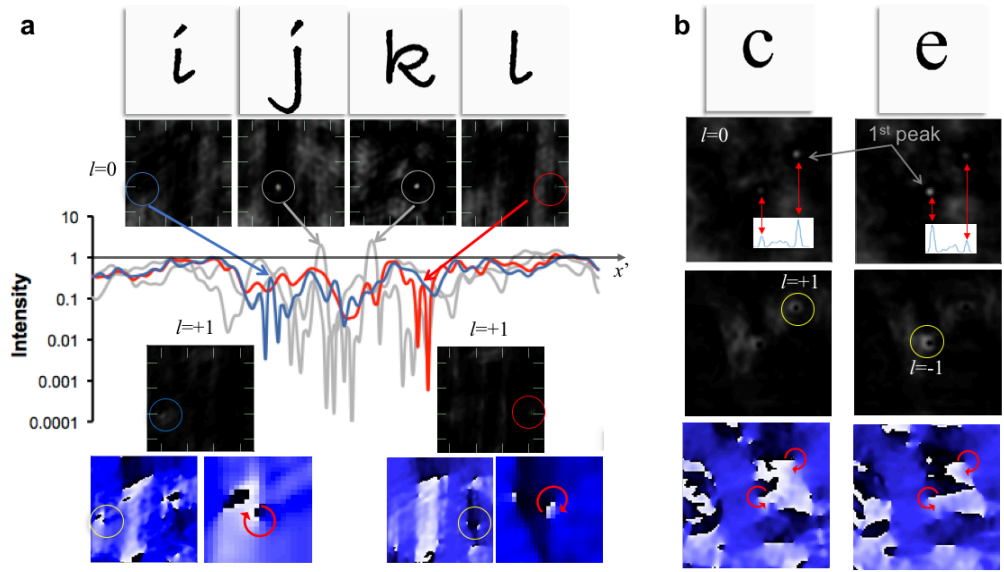

**Figure 5. False negatives and false positives**. (a) Using $W(u,v)$ with $N=16$ matched filters and $l=0$, yields low correlation peak intensities (red and blue circles). High correlation peak intensities (gray circles) for "𝑗" and "𝑘" are shown for comparison. The intensity profiles along $x'$-axis are shown for patterns "𝑖" (red) and "𝑙" (blue) as well as "𝑗" and "𝑘" (gray). Using $l=+1$ for both patterns yield phase singularities (yellow circles) with zoomed in phase distributions with red curved arrows indicating the direction of helicity from $-\pi$ to $+\pi$. (b) Using $W(u,v)$ with $N=9$ and $l=0$ yields higher correlation peak intensities but similarities of features for patterns produce secondary peaks. Encoding OAM states with opposing charges also yield phase singularities at the secondary peaks.

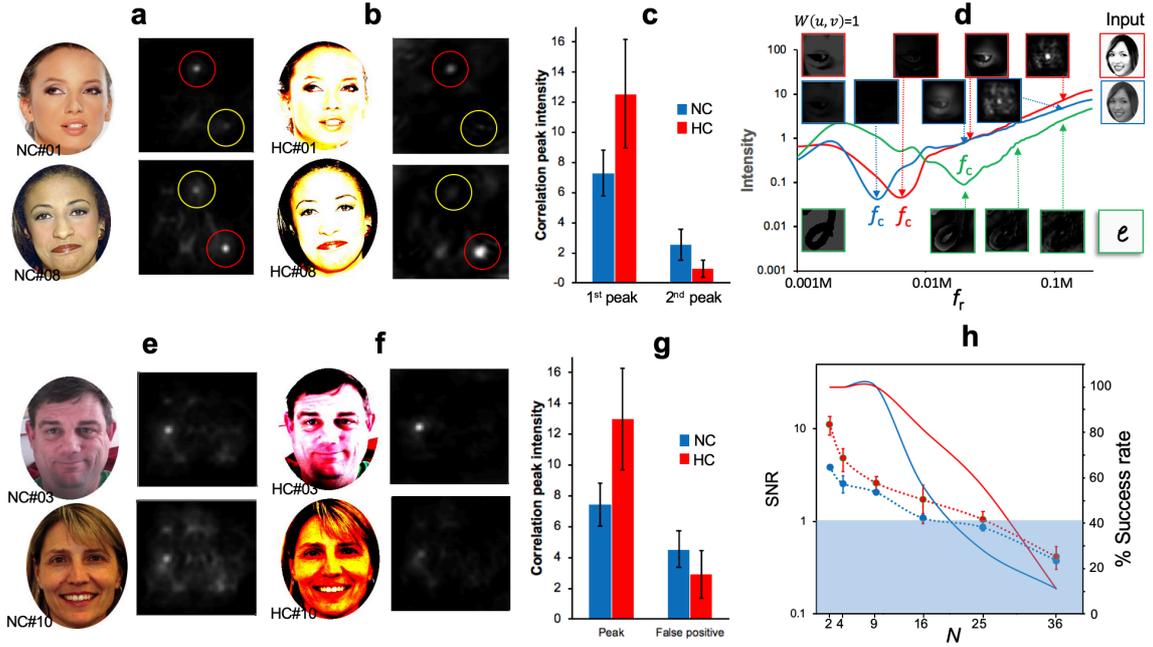

**Figure 6. Extending the spatial frequency representation in the matched filters.** (**a**) Similarities in the images within the training set yield true-positive correlation peak (red circles) and a secondary peak (yellow circles). (**b**) Changing the contrast improves the intensity ratio of the peaks. (**c**) Summary of changes in intensity ratio between normal contrast (NC) and high-contrast (HC) patterns. (**d**) Plot of the correlation peak intensity as a function of $f_r$, where $f_r$ is the radius of a circular phase-aperture at the Fourier plane. Regions greater than $f_r$ use a constant filter $W(u,v) = 1$, while regions less than $f_r$ use the filter calculated $W(u,v)$. (**e**) Similarities in the images within the training set and outside the set yield a false positive correlation peak. (**f**) Changing the contrast improves the intensity ratio of the peaks. (**g**) Summary of changes in intensity ratio between NC and HC images. (**h**) Plot of the SNR (left *x*-axis, dashed marked traces) and percentage success rate (right *y*-axis, solid unmarked traces) as a function of $N$ comparing NC (blue traces) and HC (red traces) patterns.

The correlation peaks ($l$=0) for complex multi-level patterns also yield false positives. As mentioned earlier, the bottom pattern in **Figure 4b** (NC#15), whose information was not included in the training set, produced a noticeable correlation peak like the top pattern (NC#01). Moreover, similarities between patterns within the training set also produce secondary correlation peaks (**Supplementary Figure 2a**) like those in alphanumeric patterns. **Figure 6a** shows an example where two patterns within the training set have similar features, which resulted in a 2nd peak, with a lower intensity. NC#01 shows the correlation peak (1st peak) at the top row, while the NC#08 yields its 1st peak at the bottom right. However, changing the grayscale of the face photographs to high contrast (HC) improves the correlation results (**Figure 6b, Supplementary Figure 2b**). A summary of improvement in correlation peak intensities is shown in **Figure 6c**, which includes results presented in **Supplementary Figure 2**.

**Pertinent spatial frequency information**

Differences in performance between NC and HC patterns are due to the spectral components of edges in black and white or HC patterns, which produce distinct mid-frequency phase signatures away from the optical axis. On the other hand, multi-level amplitude modulated patterns (e.g. grayscale photographs) produce low-frequency phase signatures close to the optical axis. Since the small area close to the optical axis can only multiplex a finite number of matched filters, there is a high probability for cross talk between signatures from different images resulting in secondary peaks.

To investigate the effect of the spectral components, I concatenated the spatial frequency representation to the probe their contribution for effective correlation. **Figure 6d** plots the output peak intensity for a training set with $N=9$ matched filters as a function of the radius ($f_r$) of the active area of arg$\{W(u,v)\}$. Each data point is an average of the correlation peak intensities of patterns within the training set. Representative output intensity distributions are shown for three types of inputs: HC (red), NC (blue) and alphanumeric (green) patterns. Note that each pattern type has their respective $W(u,v)$'s calculated. After training, the input patterns are expected to produce a correlation peak at specific $m$ and $n$ locations, where the output peak intensity is acquired. At $f_r=0$, the hologram at the Fourier plane is fully transmissive ($W(u,v)=1$) and the output yields a conjugate image of the input. When $f_r$ is equal to a critical frequency ($f_c$), the output produces an inverse intensity image, which sets the measured intensity at minimum. From the plot, we can deduce that for $f_r < f_c$, the spatial frequency information stored in $W(u,v)$ has minimal effect to produce an effective correlation peak. However, when $f_r > f_c$, a linear increase in the peak intensity indicates that pertinent spatial frequency information stored in $W(u,v)$ has increasingly provided effective correlation with the input patterns.

Plotting the effect of $f_r$ on the correlation peak determines the pertinent spatial frequency information necessary to be stored in the hologram. Moreover, differences in $f_c$'s indicate that the stored spatial frequency information within $W(u,v)$ is highly dependent on the pattern type. NC patterns yield an $f_c$ closest to the optical axis with the lowest spatial frequency, while black and white alphanumeric patterns have the highest $f_c$ among the three image types. HC patterns result in a slightly higher $f_c$ (compared to NC) thereby increasing the active area of stored phase signatures. Hence, the correlation of HC patterns is more effective than NC patterns.

Improving the storage of relevant phase signatures can also solve false positives for input patterns outside the training set. **Figure 6e** shows an example where a false positive occurs with an input pattern NC#10, which shows a correlation peak for NC#03. The training set contains $N=9$ matched filters for patterns NC#00 to NC#08 (**Supplementary Figure 2a**). Changing the contrast effectively discriminates and reduces the correlation peak when the photonic neuron is presented with HC#10 (**Figure 6f** and **Supplementary Figure 2b**). A summary of the changes in correlation peak intensity for true and false positives is shown in **Figure 6g**, which includes results presented in **Supplementary Figure 2**. **Figure 6h** shows the SNR (left $y$-axis, marked traces) and identification success rate (right $y$-axis, solid unmarked traces) for NC (blue trace) and HC (red trace) patterns as a function of $N$ matched filters. Hence, increasing the contrast of the visual data allows for an increased storage of relevant phase signatures for effective identification and discrimination.

### III. Discussion

The holographic photonic neuron operates as a Vander Lugt optical correlator where matched filters containing spatial frequency information of an array of visual data is stored as a phase hologram. By linking the matched filters with OAM states and multiplexing them into a single hologram, the photonic neuron can identify patterns within the training set and discriminate those outside. To allow for interconnectivity with other computing platforms, output photons carrying OAM states can be used as transmission protocol or multi-level qudits for quantum computing.

The optical correlator acts like a neuron, where the synaptic strengths are stored as pixels in a computer-generated phase-hologram. While the actual operation of real neurons is still not totally understood, the plasticity of synapses occurring within neurons provides a neurochemical foundation behind learning and memory[13]. Moreover, real neurons fire an output, which highly depends on the spatio-temporal organization as well as the strengths of the synaptic inputs[46]. Similarly, the holographic photonic neuron stores synaptic weights within each pixel in the hologram with strengths from -π to +π. Each pixel represents pertinent information from $N$ stored visual data. Here, the output of the holographic photonic neuron is a correlation peak, which is highly dependent on how the input interacts with the values of each pixel stored in the hologram.

While linking OAM with the output photons can circumvent false negatives, the holographic photonic neuron suffers from false positives, which is a common problem in in machine vision. Improving the contrast of complex multi-level (grayscale) patterns such as face photographs can improve the performance. However, the problem still exists for simple yet high-contrast patterns (e.g. alphanumeric). Similarities in the patterns produce unwanted correlation peaks that reduce the efficacy of the holographic photonic neuron. Nonetheless, the results presented here are proof-of-principle attempts to demonstrate encoding OAM states and linearly combined to form a multiplexed matched filter. In practice, multiplexed matched filters may not store entire images (faces or alphanumeric) but decomposed fragments, where the output represents only a certain characteristic feature. By identifying the critical frequency, $f_c$, of image types, fragments of pertinent spatial frequency information can be stored appropriately onto the hologram. Within a tolerable output noise level, concatenated storage is possible where the hologram can be encoded in a smaller array of phase-shifting pixels provided the encoded radial frequencies exceed $f_c$.

The holographic photonic neuron is designed as a fundamental processing unit. We can take our queue from the neural basis of the mammalian visual system. The Hubel and Wiesel [14] model of vision is hierarchical where complex visual responses build up from simpler neuronal stimuli. Moreover, multiple neurons in the visual cortex respond to certain spatial frequencies of patterns[15]. The visual cortex consists of channels of neurons that are tuned to different spatial frequencies, and their collective response results in the visual perception of patterns. Similarly, scaling and interconnecting multiple holographic photonic neurons with concatenated spatial frequency information is necessary to provide more accurate identification of the input patterns that is robust to scaling and rotation. And by transmission of OAM states, we can effectively transmit information that is independent of intensity and interconnect multiple photonic neurons to build up an effective neuromorphic machine vision processor.

Alternative machine vision approaches use of deep-learning with artificial feed-forward ANNs [28,29,47]. ANNs are used within a convolutional neural network (CNN), which operate by decoding the spatial dependencies of pixel information in a pattern. Prior to feeding the information to a feed-forward ANN, input patterns undergo a convolution with pre-set filters to find dependencies between neighboring pixels and spatial frequency information of the pattern. The convolution layer effectively finds the necessary spatial frequency signatures, which are fed into the ANN for learning and eventually identification or classification. Likewise, the holographic photonic neuron could represent different stages of the CNN where the spatial frequencies of the input pattern are decomposed [48,49]. False-positives may occur in earlier stages but will eventually be decided upon by further processing. The potential for an all-optical neuromorphic implementation of the CNN can be realized using OAM states as information carriers between photonic neurons.

## IV. Conclusion

Classical optical correlators represent certain functions in our brain, specifically in neuronal circuits responsible for our vision and our ability to recognize and classify patterns. Oftentimes, we need visual representation of the complex problem to analyze and solve accordingly. Our vision-aided analytical skills and creativity, such as making sense of mathematical equations, graphs and puzzles or writing/reading musical notes, are common examples of human intelligence that has not yet been replicated by machines. Hence, building an intelligent computer requires such skills to be integrated with deterministic computing algorithms. The future of AI hinges on designing fundamental processing units with efficient processing, concatenated localized storage as well as seamless transfer of information for scaling up to form distributed processing networks and integration between multiple computing platforms.

---

## Methods

### Numerical simulation.

The numerical simulation is performed using a program implemented on *LabView* (*National Instruments*). I used 36 alpha-numeric patterns of different fonts and 36 face photographs from the Bainbridge database [43,50]. The alphanumeric and face photographs where first converted to grayscale and centered in a $M \times M$ pixel array (where $M$=1000) prior to using them as input patterns ($a_i(x,y)$) onto a two-dimensional Fourier transform (see **Equation 1**). The 2D input patterns are set as amplitude modulated patterns with constant phase, while the phase of the output of the Fourier transform is used as matched filter. Carrier phase functions were derived using **Equation 2**, where the indices *m* and *n* can be set to identify specific locations of the correlation peak at the output plane, while *l* sets the OAM state. In the program, *m* and *n* scales the *x*- and *y*-prism functions, respectively, to shift the correlation peaks along transverse direction at the output plane, while *l* scales a helical phase around the optical axis or along the azimuthal angle φ. To derive the *N*-multiplexed matched filters, **Equation 3** is implemented via for-loop to sum the field containing the subtracted phases of the matched filter from the carrier phase function.

To implement the cross-correlation of the input patterns with the *N*-multiplexed matched filters ($W(u,v)$), the optical system was simulated to use a finite operating region around the optical axis. Hence, only a portion of the $M \times M$ array is used, which is truncated by a circular aperture of radius set to *M*/4. The relatively small aperture with respect to *M* results in a "diffraction-limited" transverse Airy pattern as correlation peaks (when *l*=0), which effectively simulates experimental conditions[51,52]. Moreover, since only the phase of $W(u,v)$ is required, the amplitude was set to unity and multiplied with the field of the Fourier transform of the input pattern. The output of the photonic neuron takes the inverse Fourier transform after the multiplication of the fields. The intensity and phase distributions of the output field are shown to characterize the output of the photonic neuron for different input patterns and OAM conditions as shown in **Figures 2**, and **4** to **6**. **Figure 6d** was plotted by varying the phase of $W(u,v)$ via a circular "phase" aperture from 0 to fully open at the Fourier plane.

**Experiment**

A benchtop experimental setup makes use of a phase-only SLM (Hamamatsu LCOS X10468-01 800 × 600) and a green laser (532nm Finesse, Laser Quantum) as shown in **Figure 3a**. Two convex lenses (*f*=200 mm) build up a 4*f* lens imaging setup where the SLM is placed at the Fourier plane. To ensure proper alignment and scaling of the calculated spectral components with experimental conditions at the Fourier plane, the system was first calibrated using the Generalized Phase Contrast (GPC) method[53].

To align the system using the GPC method, a circular aperture was placed at the input and a circular phase filter was encoded on the SLM to shift the focused zero-order beam by $\pi$ at the Fourier plane (**Supplementary Figure 3a**). Since the diameter of the input aperture sets the intensity distribution of the zero-order beam, its interaction with a circular phase filter sets a characteristic intensity distribution at the output of the 4*f* imaging setup[54,55]. When aligned properly, the filter changes the intensity distribution at the output of the 4*f* imaging setup depending on the amount of light shifted by the filter (**Supplementary Figure 3b**, bottom). When the SLM is off, the camera acquires the image of the input, and for this case, it shows an image of the laser-illuminated input aperture (**Supplementary Figure 3b**, top).

I then used the GPC method to calibrate the spectral resolution and encode the calculated matched filters arg{$W(u,v)$} onto the SLM. To calibrate, I varied the matrix size *M* to yield a similar image as the experiment. I used the image of the aperture as input and embedded it at the center of an $M \times M$ array (**Supplementary Figure 3b**, top). I then simulated the GPC method via a pair of $M \times M$ 2D Fourier transforms with variable *M* and using the same filter size as encoded onto the SLM for alignment. The image of the aperture was kept constant as *M* was varied. The intensity distribution at the center of the Fourier plane is influenced by the size of the aperture (image) with respect to the size of the array (**Supplementary Figure 3c**). Using *M*=1500 yields a closest image with the GPC experiment (**Supplementary Figure 3b**, bottom) and was hence used to calculate for $W(u,v)$ (**Supplementary Figure 3d**).

To demonstrate the performance of the holographic photonic neuron, the input patterns where projected by placing a transparency printed with alphanumeric patterns. To calculate the $W(u,v)$, the input patterns were first imaged using a camera. The face photographs were converted to grayscale and embedded/centered on an $M \times M$ array, where *M*=1500. During training, certain patterns were selected to be part of $W(u,v)$ and 2D Fourier transforms of their respective $M \times M$ arrays were calculated and combined via linear combination as described in **Equation 3**. The resulting arg{$W(u,v)$} is cropped to 600 × 600 and encoded onto the SLM (**Supplementary Figure 3a**). When the SLM is turned off, it functions as a mirror and the output registers an image of the input pattern. When the SLM is turned on and encoded with arg{$W(u,v)$}, the output registers a correlation peak when the input pattern is part of the training set. Since the calibration procedure was only a first-order approximation, the value of *M* was later fine-tuned (*M*=1480) to obtain optimum correlation peaks.

**Acknowledgements**

I thank Hans-A Bachor, Mary Jacquiline Romero and Dragomir Neshev for their insights and helpful discussions.

***

# Supplementary Figures

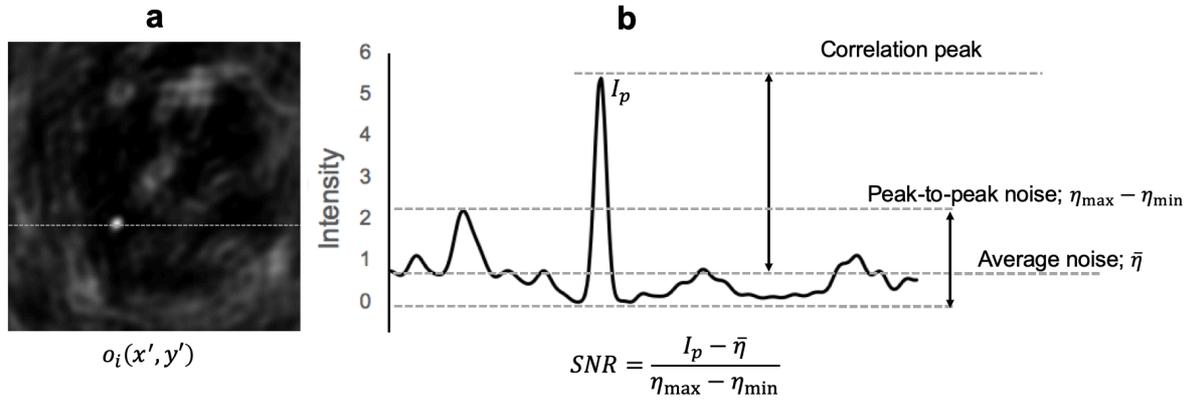

**Supplementary Figure 1. Signal-to-noise ratio for correlation output with *l*=0.** (**a**) Output 2D intensity distribution of the holographic photonic neuron using a *N*=9 multiplexed matched filter. The correlation peak is a Gaussian distribution with *l*=0. (**b**) The intensity distribution along the *x*-axis showing the noise distribution with respect to a correlation peak. The signal-to-noise ratio (SNR) is calculated using the equation. The peak ($\eta_{max}$), minimum ($\eta_{max}$) and the average noise ($\bar{\eta}$) is taken over the entire 2D intensity distribution.

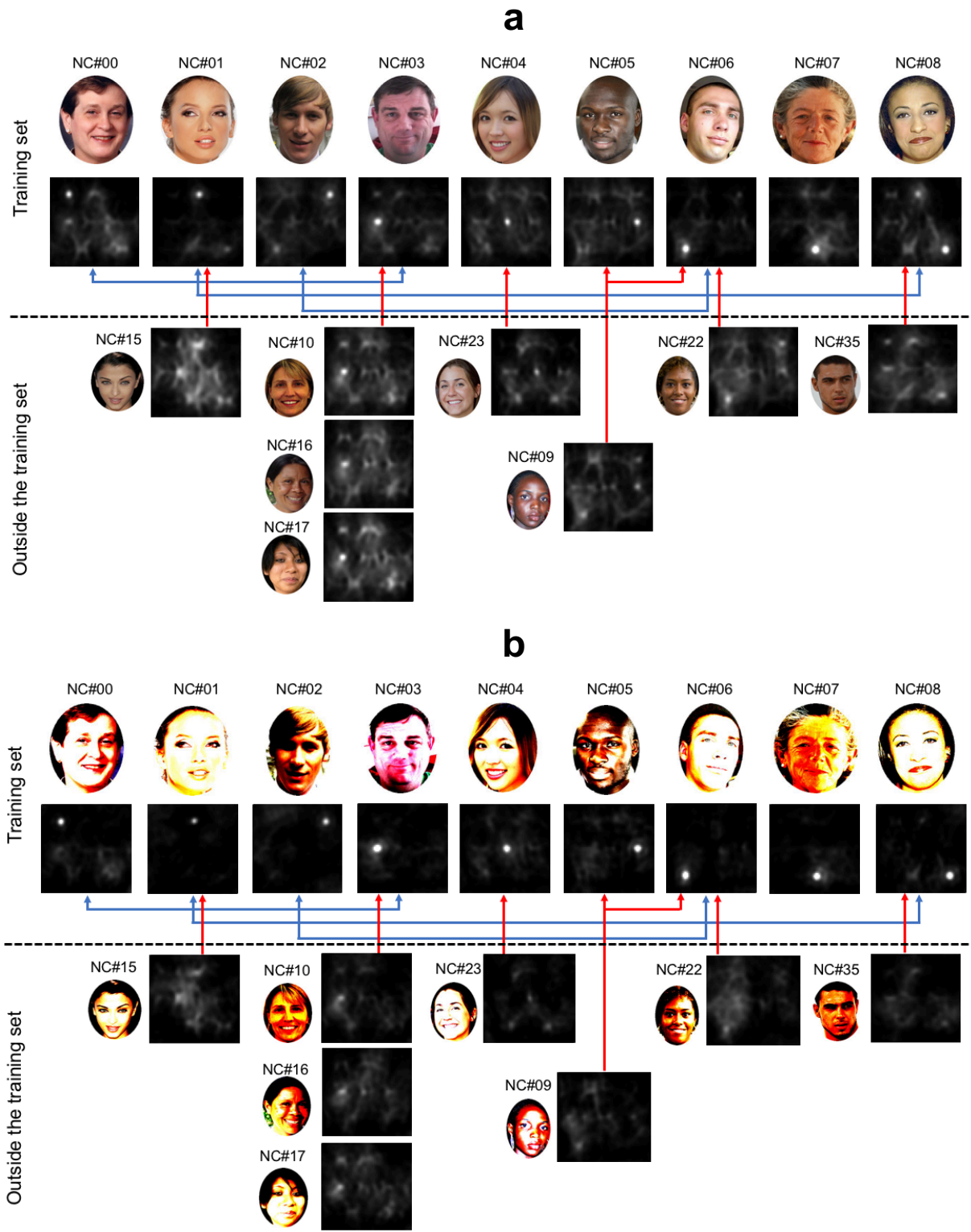

**Supplementary Figure 2. False positives in normal and high contrast patterns.** The row above the dashed line represent (**a**) normal contrast (NC) and (**b**) high contrast (HC) face photographs within the training set to derive an $N=9$ multiplexed matched filters. Similarities within the patterns in the training set yield secondary peaks. Pattern pairs where secondary peaks occur are linked by the blue lines above the black dashed line. Below the dashed line shows false positives when a correlation peak occurs when patterns outside the training set have similar features as patterns within the training set linked by red lines.

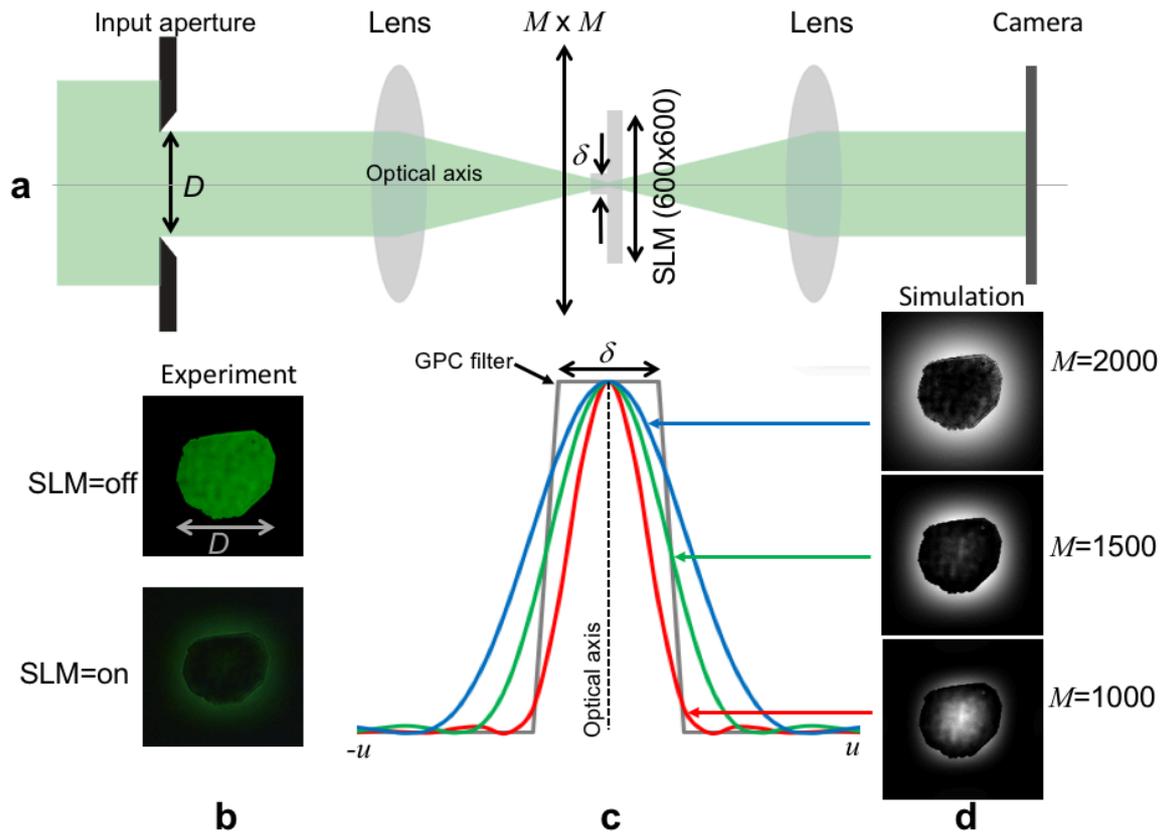

**Supplementary Figure 3. Calibrating the holographic photonic neuron using the Generalized Phase Contrast (GPC) Method**. (**a**) The GPC method built using a 4*f* imaging setup with a phase contrast filter encoded on an SLM. (**b**) Output image of the input aperture via the GPC method when the SLM is turned off (top) and turned on (bottom). (**c**) Numerically calculated intensity distribution at the Fourier plane for different $M \times M$ settings, where $M=1000$ (red), $M=1500$ (green) and $M=2000$ (blue). The lateral dimension of the phase contrast filter (gray) is shown relative to the different intensity distributions. (**d**). Numerically calculated 2D intensity distribution at the output of the GPC method for different $M \times M$ settings as indicated.